\documentclass[aps,physrev,reprint,superscriptaddress,nofootinbib,longbibliography,noeprint]{revtex4-2}

\usepackage{amsmath,amsfonts,amssymb,physics,graphicx,xcolor,soul}
\graphicspath{ {./Images/} }
\usepackage[colorlinks=true,
hyperfootnotes=true,
breaklinks=true,
citecolor=blue,
urlcolor=blue,
linkcolor=blue
]{hyperref}

\begin{document}

\title{Optimal control of a dissipative micromaser quantum battery in the ultrastrong coupling regime}

\author{Maristella Crotti}
\affiliation{Center for Nonlinear and Complex Systems, Dipartimento di Scienza e Alta Tecnologia, Università degli Studi dell'Insubria, Via Valleggio 11, 22100 Como, Italy}
\affiliation{INFN, Sezione di Milano, 20133 Milano, Italy}

\author{Luca Razzoli}
\email{luca.razzoli@unipv.it}
\affiliation{Dipartimento di Fisica ‘Alessandro Volta’, Università di Pavia, Via Bassi 6, 27100 Pavia, Italy}
\affiliation{INFN, Sezione di Pavia, 27100 Pavia, Italy}

\author{Luigi Giannelli}
\affiliation{Dipartimento di Fisica e Astronomia ‘Ettore Majorana’, Università di Catania, Via S. Sofia 64, 95123 Catania, Italy}
\affiliation{INFN, Sezione di Catania, 95123 Catania, Italy}

\author{Giuseppe A. Falci}
\affiliation{Dipartimento di Fisica e Astronomia ‘Ettore Majorana’, Università di Catania, Via S. Sofia 64, 95123 Catania, Italy}
\affiliation{INFN, Sezione di Catania, 95123 Catania, Italy}

\author{Giuliano Benenti}
\email{giuliano.benenti@uninsubria.it}
\affiliation{Center for Nonlinear and Complex Systems, Dipartimento di Scienza e Alta Tecnologia, Università degli Studi dell'Insubria, Via Valleggio 11, 22100 Como, Italy}
\affiliation{INFN, Sezione di Milano, 20133 Milano, Italy}

\date{\today}

\begin{abstract}
We investigate the open-system dynamics of a micromaser quantum battery in the ultrastrong-coupling (USC) regime. The battery consists of a quantized harmonic mode sequentially interacting, via the Rabi Hamiltonian, with a stream of qubits acting as chargers. USC enhances the charging speed but also induces unbounded energy growth and highly mixed cavity states. Dissipation suppresses this behavior, driving the system to a steady state with finite energy and ergotropy. Using optimal control theory, we show that the interplay between USC and dissipation enhances both charging performance and long-term stability against losses.
\end{abstract}

\maketitle

\section{Introduction}
Advances in quantum science and technology \cite{BenentiCasati,Acín_2018,Laucht_2021,Campbell_2025} have 
inspired investigations into energy management at the quantum scale. The concept of quantum battery---a device which stores finite amounts of energy and releases it on demand---was introduced~\cite{Alicki2013,Colloquium,Ferraro2026natrevphys}. Such devices exploit quantum hallmarks, such as coherence and entanglement, which can enhance energy-storage performance beyond classical limits~\cite{Campaioli17, Ferraro18, Rossini2019, JuliaFerre2020, Gyhm22, Rinaldi24, Andolina25, Cavaliere2025,Fischer2024avsq}.

A physically realistic and widely studied setup for implementing a quantum battery is the micromaser architecture \cite{PhysRevLett.54.551, Filipowicz:86, Meystre2007}.
It can be understood as a collision model \cite{ciccarello}, where a single-mode quantized harmonic field---such as an electromagnetic cavity mode---acts as the energy-storage unit charged by the sequential interaction with a stream of two-level natural or artificial atoms~\cite{shaghaghi2022, Salvia2023, shaghaghi2023, Massa2025}. 
In realistic implementations~\cite{Quach22,dou2023pra,Hymas2026, Joshi22,Hu22,gemme2022ibm,gemme23,Razzoli_2025,Elyasi2025,DiBello2025qst}, quantum batteries inevitably interact with their environment, dissipation and decoherence strongly affecting their performance. 
To mitigate these effects, control strategies, first applied to closed systems \cite{dou2020epl,Rodriguez2023,mazzoncini2023pra, Erdman2024,Evangelakos2024,Evangelakos2025}, have been recently extended to open quantum batteries \cite{santos2019pre,Rodriguez2024,sun2025njp}. 

In this paper, we study a micromaser operating in the ultrastrong light–matter coupling regime (USC) \cite{qin2024physrep,Kockum2019,forndiaz2019RevModPhys,Ciuti2005,Anappara2009,FornDiaz2010,Niemczyk2010,Blais2021,yoshihara2017superconducting,Hovhannisyan20,Crescente20,RidolfoEPJST2021ProbingUltrastrongLight,Dou22,GiannelliINCC2022OptimizedStateTransfer,GiannelliEPJST2023IntegratedConversionPhotodetection,GiannelliPRR2024DetectingVirtualPhotons,Crescente24,Cavaliere25}, where the interaction strength becomes comparable to the natural frequencies of the system and can no longer be treated perturbatively. In this regime, strong light–matter entanglement gives rise to complex dynamics, such as the generation of pairs of excitations reminiscent of the dynamical Casimir effect. This regime is expected to significantly enhance the battery performance, enabling faster charging via interaction channels, irrelevant at weaker coupling, which may even lead to unbound energy growth.

A realistic scenario emerges treating the micromaser battery as an open quantum system~\cite{Manzano,Campaioli,breuer2002theory,vacchini2024open,Beaudoin2011,weiss-book}, the battery and the chargers forming the ``principal'' system which interacts with an environment of uncontrolled degrees of freedom. Dissipation naturally stabilizes the battery, preventing unbounded energy growth.
We then exploit optimal control techniques \cite{Koch2022,Giannelli2022} to design protocols that enhance the ergotropy and stabilize it against dissipative losses via an optimized measurement-based passive-feedback strategy.
Despite extensive studies on quantum batteries, the combined effects of USC and dissipation remain largely unexplored. 
Our work addresses this gap by highlighting the key role of their interplay, thereby helping to clarify the limits and opportunities of micromaser quantum batteries.

The paper is organized as follows. In section~\ref{sec:model}, we introduce the model, the master equation governing the open system dynamics, and the figures of merit by which we characterize the battery's performance. Then, after discussing the results for both the non-dissipative and dissipative system dynamics, in section~\ref{sec:num_results} we elaborate on optimal control strategies for the charging and stabilization protocols. Finally, in section~\ref{sec:conclusion} we present our main conclusions and perspectives.

\begin{figure}[!t]
    \centering
    \includegraphics[width=\columnwidth]{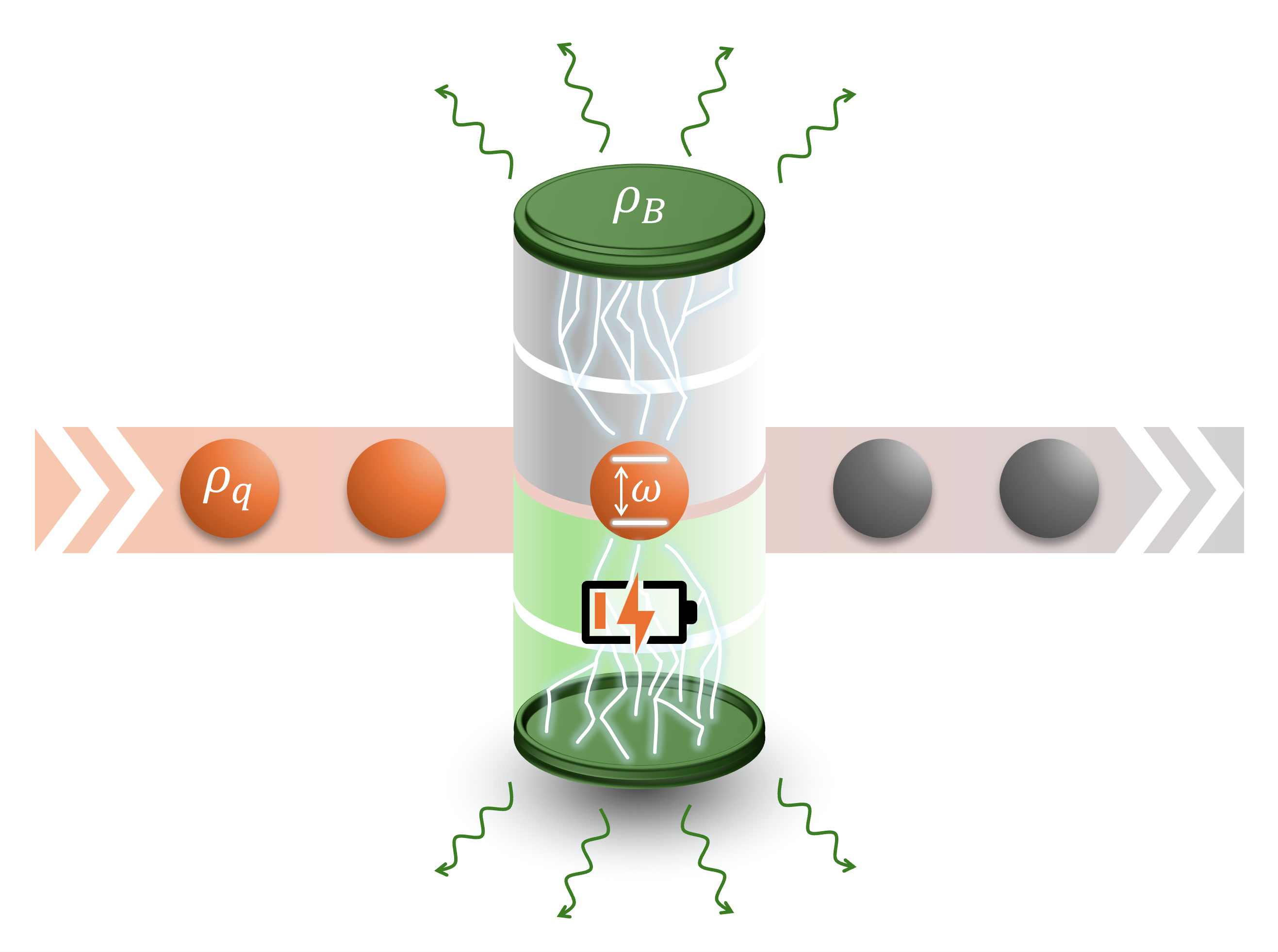}
    \caption{
    Pictorial representation of a micromaser quantum battery. The battery, modeled as a single-mode cavity, interacts sequentially with a stream of identical qubits. Each qubit enters the cavity, transfers energy to the field, and is discarded upon exit. Wavy arrows denote cavity losses. The interaction time and qubit preparation are optimized to maximize energy transfer to the cavity field.}
    \label{sketch_model}
\end{figure}

\section{Model}
\label{sec:model}

\subsection{Micromaser quantum battery} 
\label{sec:micromaser_qb}
The focus of our analysis is a specific and conceptually rich model of quantum battery based on the micromaser architecture. 
The micromaser battery consists of a single-mode electromagnetic field confined in a high-quality cavity, which serves as the quantum battery ($B$), that sequentially interacts with a stream of identically prepared two-level quantum systems (atomic~\cite{Harochebook} or superconducting~\cite{Blais2021} qubits) $\{q_k\}$ \cite{shaghaghi2022, shaghaghi2023} (see figure \ref{sketch_model} for a pictorial representation). 
In the ideal model, the stream of qubits has fixed velocity and the rate of injection is low enough that the atoms 
traverse the cavity one by one (i.e., there are never two atoms in the cavity at the same time).
Moreover, under realistic conditions, qubits can be assumed to be non-interacting and initially uncorrelated with each other.
Therefore, the dynamics is described by a basic Markovian collision model.

The interaction between the qubit and the cavity mode is well described by the quantum Rabi model \cite{Meystre2007,braak2011prl}, whose total Hamiltonian is given by\footnote{
In what follows, we discuss the micromaser quantum battery in the Schr{\"o}dinger picture, where the total Hamiltonian~\eqref{eq:H_tot_micromaser} is time-independent. This choice allows us to directly apply the GKLS master equation~\eqref{eq:GKLS_me}. In the interaction picture, instead, the presence of counter-rotating terms renders the Hamiltonian~\eqref{eq:H_tot_micromaser} explicitly time-dependent, thereby complicating both the derivation and the interpretation of the open system dynamics. We stress this point because previous works addressed the micromaser battery dynamics in the interaction picture \cite{shaghaghi2022,shaghaghi2023}
However, one should keep in mind that formulating a collision model in the interaction picture or in the Schr{\"o}dinger picture generally corresponds to different physical scenarios, depending on the adopted representation \cite{razzoli2026cmspip}.}
\begin{equation}
        \hat{H}_{B,q} = \hat{H}^{(0)}_{B,q} + \hat{V}_{B,q}, 
        \label{eq:H_tot_micromaser}
    \end{equation}
where the free and interaction Hamiltonians are, respectively,
\begin{align}
      \hat{H}^{(0)}_{B,q}& \equiv \hat{H}_B + \hat{H}_{q} = \hbar\omega_B \hat{a}^\dagger \hat{a} + \hbar\frac{\omega_q}{2} \hat{\sigma}_z, \\
      \hat{V}_{B,q}& \equiv \hbar g \left( \hat{a} \hat{\sigma}_+ + \hat{a}^\dagger \hat{\sigma}_- + \hat{a}^\dagger \hat{\sigma}_+ + \hat{a} \hat{\sigma}_- \right).
\end{align}
Here, $\hat{a}^{(\dagger)}$ is the cavity annihilation (creation) operator, $\hat{\sigma}_{+(-)}$ is the qubit raising (lowering) operator, $\hat{\sigma}_{z}$ is the Pauli $z$ operator, $\hbar$ is the reduced Planck constant (we set $\hbar=1$ throughout), \( \omega_{q(B)} \) is the qubit (cavity) frequency, and $g$ is the coupling strength between the field and the qubit. 
We focus on the resonant case where the qubit and cavity frequencies are matched, i.e., \( \omega_q = \omega_B \equiv \omega \).

The coupling constant $g$ measures the strength of the interaction between the qubits and the cavity. In cavity-QED implementations~\cite{Harochebook}, when the coupling strength is much smaller than the bare frequencies ($g \ll \omega_B, \omega_q$), the dynamics is accurately captured by the Jaynes--Cummings (JC) model. 
This model is obtained by applying the rotating-wave approximation (RWA), which allows us to neglect the so-called \emph{counter-rotating terms}, the last two terms in the interaction Hamiltonian $\hat{V}_{B,q}$, that do not conserve the excitation number $\hat{\cal{N}}=\hat{a}^\dagger\hat{a}+\hat{\sigma}_+\hat{\sigma}_-$ 
of the system. 
However, in 
circuit-QED~\cite{Blais2021} one can address the USC regime, where $g$ becomes comparable to the cavity and qubit frequencies, the RWA breaks down, and the JC model is no longer valid~\cite{Kockum2019}. 
Light and matter become strongly coupled, and the counter-rotating processes significantly modify the excitation-exchange mechanisms. 
These non-perturbative features have been shown to enhance charging speed and energy-storage capabilities in quantum batteries~\cite{shaghaghi2022}, which motivates our focus on the micromaser specifically operating in the USC regime.

Each qubit is initialized in a partially coherent state
\begin{align}
  \label{eq:qubit_state}
    \rho_q = &\, q \ketbra{g}{g} + (1 - q) \ketbra{e}{e} \nonumber\\
    &+ c \sqrt{q(1 - q)} \left( \ketbra{e}{g} + \ketbra{g}{e} \right),
\end{align}
where \( \ket{g} \) and \( \ket{e} \) denote the ground and excited states of the qubit, respectively. 
The parameter \( q \in [0, 1] \) controls the initial population inversion of the qubit, while the coherence parameter \( c \in [0, 1] \) determines the purity of the initial qubit state. The battery is initialized in its ground state, i.e., \( \rho_B=\dyad{0} \).

During each interaction interval of duration \( \tau \), the total system (cavity + qubit) evolves unitarily under the Hamiltonian \( \hat{H}_{B,q} \) \eqref{eq:H_tot_micromaser}. 
The state of the battery after each collision is obtained by tracing out the qubit
\begin{equation}
  \label{eq:rho_B}
    \rho_{B}(k) = \Tr_q \left[ \hat{U}_{B,q}(\tau, 0) \left( \rho_{B}(k-1) \otimes \rho_q \right) \hat{U}^\dagger_{B,q}(\tau,0) \right],
\end{equation}
where \( \rho_{B}(k) \) is the battery state after the \( k \)th collision, and \( \hat{U}_{B,q}(t,t_0) = \exp\big[-i \hat{H}_{B,q} (t-t_0)\big] \) is the time evolution operator generated by the total Hamiltonian. 
This iterative evolution defines the overall dynamics of the micromaser battery.

After each collision, letting $\hat{H}_B=\omega\hat{a}^\dagger \hat{a}$ be the battery Hamiltonian, we monitor three relevant figures of merit, that serve as indicators of the battery's performance and coherence throughout the charging protocol:
(i) the energy stored into the cavity (measured from the ground state), \( E(k) = \Tr\big[ \hat{H}_B \rho_{B}(k) \big] \);
(ii) the purity of the cavity state, \( \mathcal{P}(k) = \Tr\big[ \left( \rho_{B}(k) \right)^2 \big] \), which we use as an indicator of the degradation of the battery state;
and (iii) the \emph{ergotropy} $\mathcal{E}(k)$ stored into the cavity, which quantifies the maximum amount of work extractable from the battery state $\rho_B(k)$ with respect to the Hamiltonian $\hat{H}_B$ via unitary transformations.
Ergotropy is the appropriate figure of merit for globally extractable work; however, when a system comprises interacting subsystems and work is extracted locally, the ergotropy of a subsystem does not in general capture the maximum extractable work, and one should instead resort to the (extended) local ergotropy \cite{salvia2023pra,castellano2024prl} to account for Hamiltonian couplings. In our protocol, although such considerations are in principle relevant---especially in the USC regime---we assume that work extraction occurs only after the battery is decoupled from the qubit (charger). 
Therefore, the standard ergotropy of the battery provides the appropriate figure of merit.

For a quantum system with Hamiltonian $\hat{H}=\sum_{k} \epsilon_k \dyad{\epsilon_k}$ with $\epsilon_{k+1} \geq \epsilon_{k}$ prepared in a state $\rho = \sum_{k} r_k \ketbra{r_k}{r_k}$ with $r_{k+1} \leq r_k$, the ergotropy is defined as \cite{Allahverdyan2004}
\begin{equation}
      \label{eq:ergotropy_definition}
      \mathcal{E}(\rho) \equiv \Tr\big[\hat{H}\rho\big] - \min_{\hat{U}} \Tr\big[\hat{H} \hat{U}\rho \hat{U}^\dagger\big],
\end{equation}
where the minimization is performed over all unitary operators $\hat{U}$.
Solution of the minimization problem is the unitary $\hat{U}_{\mathcal{E}}=\sum_{k} \ketbra{\epsilon_k}{r_k}$, which maps the initial state $\rho$ into $\pi_\rho \equiv \hat{U}_{\mathcal{E}} \rho \hat{U}_{\mathcal{E}}^\dagger = \sum_{k} r_k \ketbra{\epsilon_k}{\epsilon_k}$. The state $\pi_\rho$ is referred to as \textit{passive}, as it represents a quantum state from which no work can be extracted through unitary operations. In general, it can be shown that a state is passive if and only if it is diagonal in the energy eigenbasis $\{\vert \epsilon_k \rangle \}$ and its eigenvalues decrease monotonically with energy (i.e., there is no population inversion) \cite{Pusz1978,Lenard1978},
  \begin{equation}
      \label{eq:passive_state_theorem}
      \pi = \sum_{k} p_k \ketbra{\epsilon_k}{\epsilon_k}, \quad p_{k+1} \leq p_k~.
  \end{equation}
Notably, provided the Hamiltonian $\hat{H}$ has a non-degenerate spectrum,\footnote{If the Hamiltonian spectrum is degenerate, the passive state
is defined up to unitaries acting in each degenerate
subspace.} for any state $\rho$ there exists a unique passive state $\pi_\rho$, and the ergotropy can be expressed in terms of such state
$\mathcal{E}(\rho) = \Tr\big[\hat{H}\rho\big]-\Tr\big[\hat{H}\pi_\rho\big]$.

\subsection{Dissipation}

So far, we have described an ideal model, treating the micromaser battery as a closed quantum system. 
In practice, however, the system is never completely isolated from its environment. Interactions with the surroundings introduce dissipation and decoherence, which can strongly affect the system's dynamics. 

To account for these effects, we describe the battery using the open-quantum-system framework developed in Ref.~\cite{Beaudoin2011}.
We model dissipation occurring during each qubit--cavity interaction, rather than only during idle periods between collisions as in previous studies~\cite{shaghaghi2023}. 
To compensate for energy loss due to dissipation, qubits are injected in immediate succession, keeping the cavity continuously engaged in the charging process.
This assumption leads to a more realistic model of the micromaser battery, as it accounts for the fact that the environment remains present and active even during the charging process. 
As a result, the dissipative dynamics should act on the full system composed of the cavity and the interacting qubit, which we denote collectively as \( S \). 

We model the environment as a thermal bath of harmonic oscillators in thermal equilibrium at inverse temperature $\beta = 1/(k_B T)$ (we set the Boltzmann constant $k_B=1$ throughout), with Hamiltonian  
\begin{equation}
    \hat{H}_E = \sum_k \omega_k \hat{b}_k^\dagger \hat{b}_k,
\end{equation}
where \( \hat{b}_k \) and \( \hat{b}_k^\dagger \) are the annihilation and creation operators for the \( k \)th bath mode of frequency \( \omega_k \). 
The system--bath interaction is described by a coupling of the form
\begin{equation}
    \hat{H}_{I} = \hat{V} \otimes \sum_k g_k (\hat{b}_k^\dagger + \hat{b}_k),
\end{equation}
where \( \hat{V} \) is the system operator through which \( S \) couples to the bath, and \( g_k \) are coupling constants.

Following the standard theory of open quantum systems~\cite{Manzano,Campaioli,breuer2002theory,vacchini2024open}, the evolution of the joint system state \( \rho_S(t) \) is governed by a GKLS master equation \cite{kossakowski1972nonhamiltonian, lindblad1976generators, gorini1976completely}
\begin{align}
    &\frac{d}{dt} \rho_S(t) = -i [\hat{H}_S, \rho_S(t)] \nonumber\\
    &+ \sum_{\omega} \left( \hat{L}_\omega \rho_S(t) \hat{L}_\omega^\dagger - \frac{1}{2} \left\{ \hat{L}_\omega^\dagger \hat{L}_\omega, \rho_S(t) \right\} \right),
    \label{eq:GKLS_me}
\end{align}
where \( \hat{H}_S = \sum_\epsilon \epsilon \ketbra{\epsilon}{\epsilon}\)
is the total Hamiltonian of the qubit--cavity system during the collision (see equation \eqref{eq:H_tot_micromaser}), $[\cdot,\cdot]$ denotes the commutator, and $\{\cdot,\cdot\}$ the anticommutator.
Here, the sum runs over the Bohr frequencies \(\omega = \epsilon' - \epsilon\), which correspond to energy differences between eigenstates \(|\epsilon'\rangle\) and \(|\epsilon\rangle\) of \(\hat{H}_S\). These frequencies determine the allowed transitions through which the system can exchange energy with the environment.
We stress that the GKLS master equation \eqref{eq:GKLS_me} governs the dissipative dynamics of the qubit–cavity system during each single collision, and is not obtained as an effective continuous-time description of the cavity dynamics in the limit of infinitesimal collision times

The Lindblad operators \( \hat{L}_\omega \) describe energy exchanges between the system and the environment, and are defined as
\begin{equation}
    \hat{L}_\omega = \sqrt{\gamma(\omega)} \sum_{\epsilon' - \epsilon = \omega} V_{\epsilon \epsilon'} |\epsilon\rangle \langle \epsilon' |,
\end{equation}
where
\( V_{\epsilon \epsilon'} = \langle \epsilon | \hat{V} | \epsilon' \rangle \). 
Although the dissipation acts on the full system \( S \), we assume that only the cavity is directly coupled to the environment. Accordingly, we take the system operator \( \hat{V} \) to be
\begin{equation}
    \hat{V} = (\hat{a} + \hat{a}^\dagger) \otimes \mathbb{I}_q,
\end{equation}
where \( \mathbb{I}_q \) is the identity on the qubit subspace. 
This reflects the physical picture in which the cavity is a fixed, central component of the micromaser battery, permanently present and exposed to the environment, whereas the qubits are transient ancillae injected one at a time.         
    
The decay rate associated with a transition at Bohr frequency \( \omega \) is given by \cite{Albash_2012, cenedese2025prb}
\begin{equation}
    \gamma(\omega) = \frac{2\pi J(|\omega|)}{1 - e^{-\beta |\omega|}} \left( \Theta(\omega) + e^{-\beta |\omega|} \Theta(-\omega) \right),
\end{equation}
where \( \Theta(\omega) \) is the Heaviside step function and \( J(\omega) \) is the spectral density of the environment. The two contributions describe spontaneous and stimulated emission (\( \omega > 0 \)) and absorption (\( \omega < 0 \)) processes, respectively.

We assume an Ohmic spectral density with exponential cutoff,
\begin{equation}
    J(\omega) = \eta |\omega| e^{-|\omega| / \omega_c},
\end{equation}
where \( \eta \) is a coupling constant and \( \omega_c \) is a cutoff frequency characterizing the bath's correlation timescale. 

\section{Numerical results}
\label{sec:num_results}
We present here the results of the numerical investigation of the micromaser quantum battery dynamics. 
We begin by exploring the closed-system dynamics of the battery in the absence of dissipation, then turn to the open-system case to investigate how environmental effects influence the dynamics, and finally consider parameter optimization for both the charging and stabilization processes.
All simulations are carried out using the open-source \texttt{QuTiP} library for Python~\cite{Johansson2012,Johansson2013}, while the optimization of system parameters is performed with the \texttt{minimize} routine from \texttt{scipy.optimize}, employing the limited-memory Broyden-Fletcher-Goldfarb-Shanno (L-BFGS-B) quasi-Newton method~\cite{LBFGSB1995,2020SciPyNMeth}, with gradients estimated via finite difference, to efficiently explore the parameter landscape.
This algorithm is an iterative gradient-based optimization method for nonlinear problems with simple bound constraints. It determines the search direction by combining gradient information with a limited-memory approximation of the inverse Hessian matrix, progressively updated via a quasi-Newton (secant) scheme, thus efficiently incorporating curvature information without explicitly computing second derivatives.

\begin{figure*}[!t]
    \centering
    \includegraphics[width=\linewidth]{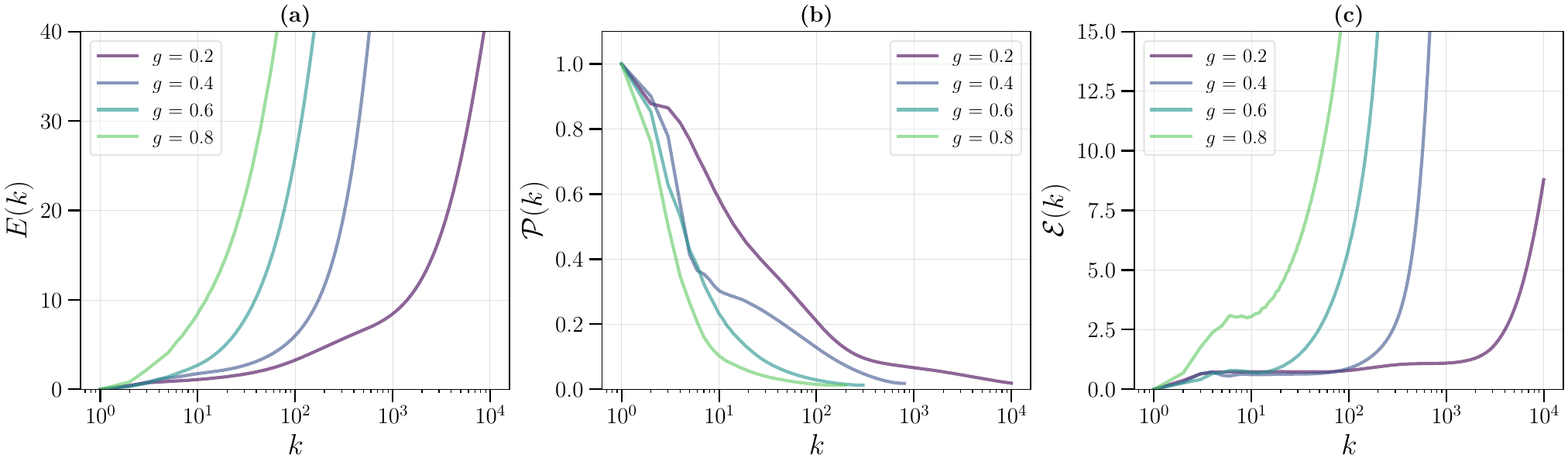}
    \caption{Evolution of the closed quantum battery: (a) energy, (b) purity, and (c) ergotropy as functions of the number of collisions. Each curve corresponds to a different coupling strength in the USC regime, $g \in \{0.2,0.4,0.6,0.8\}$, with fixed parameters $c=1$, $q=0.5$, $\tau=10$, and $\omega=1$. 
    In the closed-system regime, both the cavity energy and ergotropy increase without bound as the number of collisions grows, while the purity rapidly decreases, indicating that the battery evolves into strongly mixed states.}
    \label{fig:closed_dynamics}
\end{figure*}

\subsection{Non-dissipative dynamics}
We start our study by numerically investigating the charging dynamics of the micromaser quantum battery in the 
non-dissipative regime, where each qubit-cavity collision is treated as a unitary evolution (closed system dynamics). Although the battery's reduced dynamics are open due to repeated interactions, we refer to this as a closed quantum battery, as there is no coupling to an external environment.
Figure~\ref{fig:closed_dynamics} shows the energy, purity, and ergotropy as functions of the number of collisions, for four values of the coupling strength $g$ in the USC regime.
Both the cavity energy and ergotropy increase indefinitely with the number of collisions, showing no sign of stabilization, while the purity rapidly decreases, indicating that the battery evolves into strongly mixed states. 
These effects are more pronounced at larger coupling strengths, with faster growth of energy and ergotropy and a more rapid decrease of purity, reflecting the stronger energy transfer and enhanced mixing induced by the counter-rotating terms in the Rabi Hamiltonian \eqref{eq:H_tot_micromaser}.
Our simulations also confirm that variations in the qubit preparation have only a modest influence on this behavior: coherence plays a minor role, while population inversion mainly affects the charging rate.
 
Overall, in the absence of dissipation, the counter-rotating terms
lead to unbounded energy growth and increasingly mixed cavity states, which makes it challenging to control and exploit the energy stored in the cavity.

\begin{figure*}[!t]
    \centering
    \includegraphics[width=\linewidth]{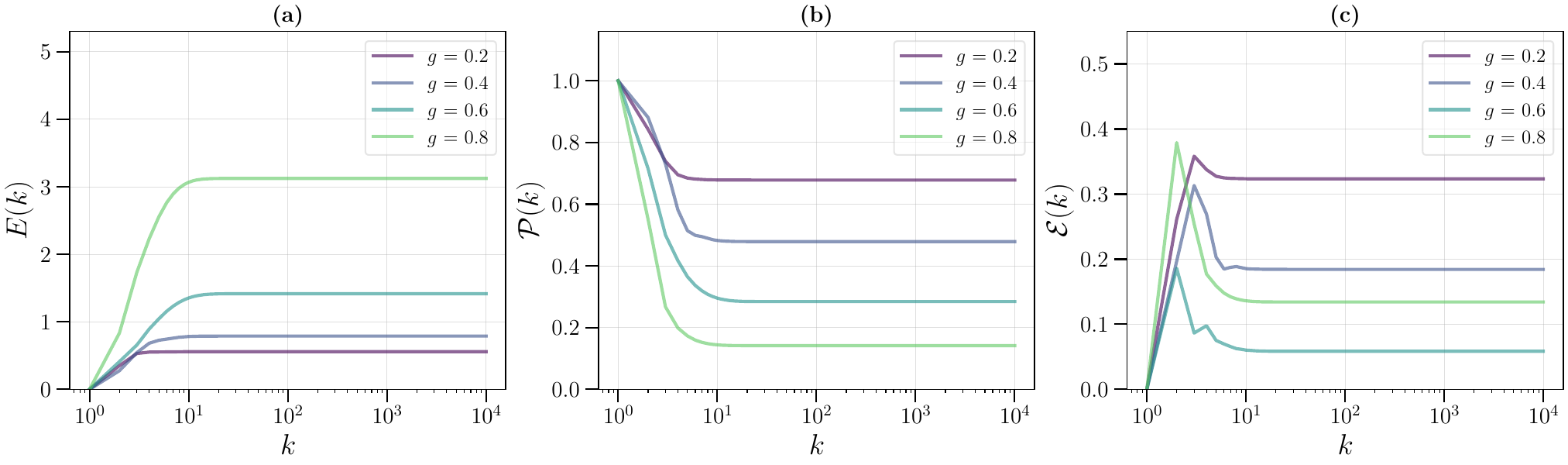}
    \caption{Evolution of the open quantum battery: (a) energy, (b) purity, and (c) ergotropy as functions of the number of collisions. Each line corresponds to a different value of $g \in \{0.2,0.4,0.6,0.8\}$ in the USC regime, with fixed system parameters $c=1$, $q=0.5$, $\tau=10$, $\omega=1$, and environment parameters $\omega_c=3$, $\beta=450$, and $\eta=0.01$. 
    Including dissipation during each qubit--cavity collision qualitatively changes the dynamics: energy and ergotropy converge to well-defined steady-state values, while purity decreases moderately with increasing $g$. The steady-state ergotropy remains nonzero, indicating that the energy stored in the cavity is at least partially extractable through unitary operations.}
    \label{fig:open_dynamics}
\end{figure*}

\subsection{Dissipative dynamics}
\label{sec:open_dynamics}
We then turn to the numerical study of the dissipative dynamics of the micromaser quantum battery, where each qubit-cavity collision is treated as the evolution of an open quantum system due to the coupling to an external environment. We therefore refer to this as an open quantum battery.
Including dissipation during each qubit-cavity collision qualitatively changes the dynamics in the USC regime.
As shown in figure \ref{fig:open_dynamics}, energy and ergotropy, which grew unboundedly in the closed system, now converge to well-defined steady-state values.
Purity is also higher than in the non-dissipative case, particularly at small coupling, and decreases moderately with increasing $g$. 
Importantly, the steady-state ergotropy is nonzero, indicating that the energy stored in the cavity remains at least partially extractable through unitary operations.
These results demonstrate that environmental dissipation provides a stabilizing mechanism, suppressing the instabilities induced by counter-rotating terms and allowing the micromaser battery to store finite, extractable energy.

In our analysis of the open system dynamics of the micromaser battery, we use environmental parameters $\eta = 0.01$, $\beta = 450$, and $\omega_c = 3$, corresponding to a dimensionless single-photon loss rate of $\gamma/\omega \simeq 0.045$, which yields a decay time of $T_1 \approx 22$ (in units of $\omega^{-1}$). In contrast, typical cavity decay rates reported for experimental setups operating in the USC regime, such as those in circuit QED \cite{Niemczyk2010, Beaudoin2011, Fink2008, Grimsmo2013}, are of the order of $10^{-4}$ (in units of the cavity frequency), i.e., about two orders of magnitude lower than the effective loss rate used in our simulations. 
The reason for employing comparatively large dissipation values, while still remaining within the weak system-environment coupling regime required by the GKLS formalism, is primarily computational. Numerically implementing the single-mode field of the cavity requires a truncated Fock basis with a sufficient number of excited states: the larger the Hilbert space, the more accurate the simulation, the higher the computational cost. In practice, one restricts the Fock basis to the minimum number of excited states such that the highest ones are not effectively involved in the dynamics. When a continuous stream of qubits interacts with the cavity, low dissipation typically allows highly excited states to become significantly populated during the charging process, saturating the truncated Fock basis, and thus compromising the reliability of the numerical simulation.
Stronger dissipation, instead, allows us to keep the cavity population well within the numerically accessible subspace, while still retaining meaningful physical behavior. Therefore, our simulations, which already reveal promising results for both ergotropy and stored energy, suggest that the performance of the micromaser battery would likely be further enhanced under realistically weaker dissipation.

\subsection{Optimal control}
The results obtained so far motivate the next step of our analysis: designing optimized charging and stabilization protocols for the dissipative micromaser quantum battery. We consider a setting motivated by realistic experimental conditions, where the total number of collisions between the qubits (chargers) and the cavity (battery) is fixed in advance due to practical constraints, and the charging protocol is optimized under this constraint.
This choice is further justified by the fact that, in general, the steady-state ergotropy can be significantly lower than the maximum ergotropy attained during the transient dynamics (see figure \ref{fig:open_dynamics}, particularly for larger values of $g$), suggesting that steady-state optimization is not necessarily the most relevant objective for energy-storage purposes.
It is therefore more appropriate to focus on the early-stage dynamics, which are most relevant for the charging process, and to formulate the problem in terms of maximizing the achievable ergotropy over a finite number of collisions together with its subsequent stabilization against dissipation.

Optimal control theory provides a systematic framework to achieve these goals, by tuning control parameters such as the qubit population inversion and the interaction times with the cavity to optimize a well-defined cost function. In the context of quantum batteries \cite{Hu22,mazzoncini2023pra}, this enables the design of charging and storage protocols that enhance the battery's performance, even in the presence of dissipation and non-ideal effects.

\subsubsection{Charging process}
\label{sec:opt_ctrl_charging}
Our goal is to design an optimal charging protocol for the micromaser quantum battery that maximizes the stored ergotropy, under the constraint of a limited amount of resources, i.e., a finite stream of qubits.
We optimize over two sets of control parameters: the initial population inversion of the qubit, $q$ (see equation~\eqref{eq:qubit_state}), which is taken to be the same for all collisions, and the interaction times, $\{\tau_k\}$, that are allowed to vary from collision to collision.
The optimization problem---i.e., maximizing the final ergotropy of the system, $\mathcal{E}_{F}$---is reformulated as the minimization of a \emph{cost function} defined as $\mathcal{C} = - \mathcal{E}_{F}$.

Assuming a realistic setting of limited resources, the charging protocol is optimized under the constraint of a fixed number of qubit-cavity collisions. For the dissipation parameters used in figure \ref{fig:open_dynamics}, the system approaches its steady state after $N_\text{col}\simeq 10$ collisions; we therefore restrict our simulations to $N_\text{col}=5$, as the early-stage collisions are the most relevant for the charging process.
We set the physical parameters of the model as follows:
$c = 1$\footnote{Initially, we also optimized the parameter $c$; however, we observed that most optimizations yielded $c=1$ as the optimum. Therefore, to improve computational efficiency, we fixed $c=1$ and excluded it from further optimization.} and $\omega=1$ for the system, and $\eta=0.01$, $\omega_c = 3$, $\beta = 450$ for the environment, as in section~\ref{sec:open_dynamics}.
While keeping these parameters fixed, we scanned values of the coupling constant $g \in [0.1, 0.7]$, ranging from weaker to stronger coupling strengths while remaining in the USC regime.

Optimization problems often suffer from local minima of the cost function, which result in sub-optimal solutions. To increase confidence that the obtained minimum of $\mathcal{C}$ corresponds to a \emph{global} minimum, rather than a local one, we performed multiple optimizations starting from different initial conditions for the parameters to be optimized, randomly sampled within preset boundaries. This approach allows us to explore different regions of the parameter space of $ q $ and $ \{\tau_k\} $, thereby increasing the likelihood that the resulting configuration indeed maximizes the final ergotropy $\mathcal{E}_F$.

To assess the effectiveness of our optimized protocol, we compare our results with a natural benchmark choice of collision times,
\begin{equation}
    \tau_k = \frac{\pi}{2 g \sqrt{k}}, \quad k = 1, \dots, N_\textrm{col},
    \label{eq:tau_pi_pulse}
\end{equation}
while fixing $q=0$, meaning the qubit is prepared in its excited state. This choice is motivated by the ideal JC model at resonance, where the interaction between a two-level system and a quantized field mode produces Rabi oscillations with frequency
\(
\Omega_n = 2 g \sqrt{n}.
\) Setting the interaction time to $\tau_n=\pi/\Omega_n$ implements a $\pi$-pulse on the transition $\ket{n-1,e} \leftrightarrow \ket{n,g}$. The optimal charging protocol in the JC model thus requires $q=0$ and $\tau_n$, as this maximizes the energy transfer at each collision by making the qubit relax to the ground state, $\ket{e} \to \ket{g}$, by transferring the excitation to the cavity, $\ket{n-1} \to \ket{n}$, therefore effectively charging the cavity by one photon per interaction.
We refer to this prescription as the $\pi$-pulse protocol in the following. This provides a natural benchmark against which we can compare the performance of our optimized protocol, even though the dynamics of our micromaser model are governed by the full Rabi Hamiltonian \eqref{eq:H_tot_micromaser} and are subject to dissipation.

\begin{figure*}[!t]
    \centering
    \includegraphics[width=0.8\linewidth]{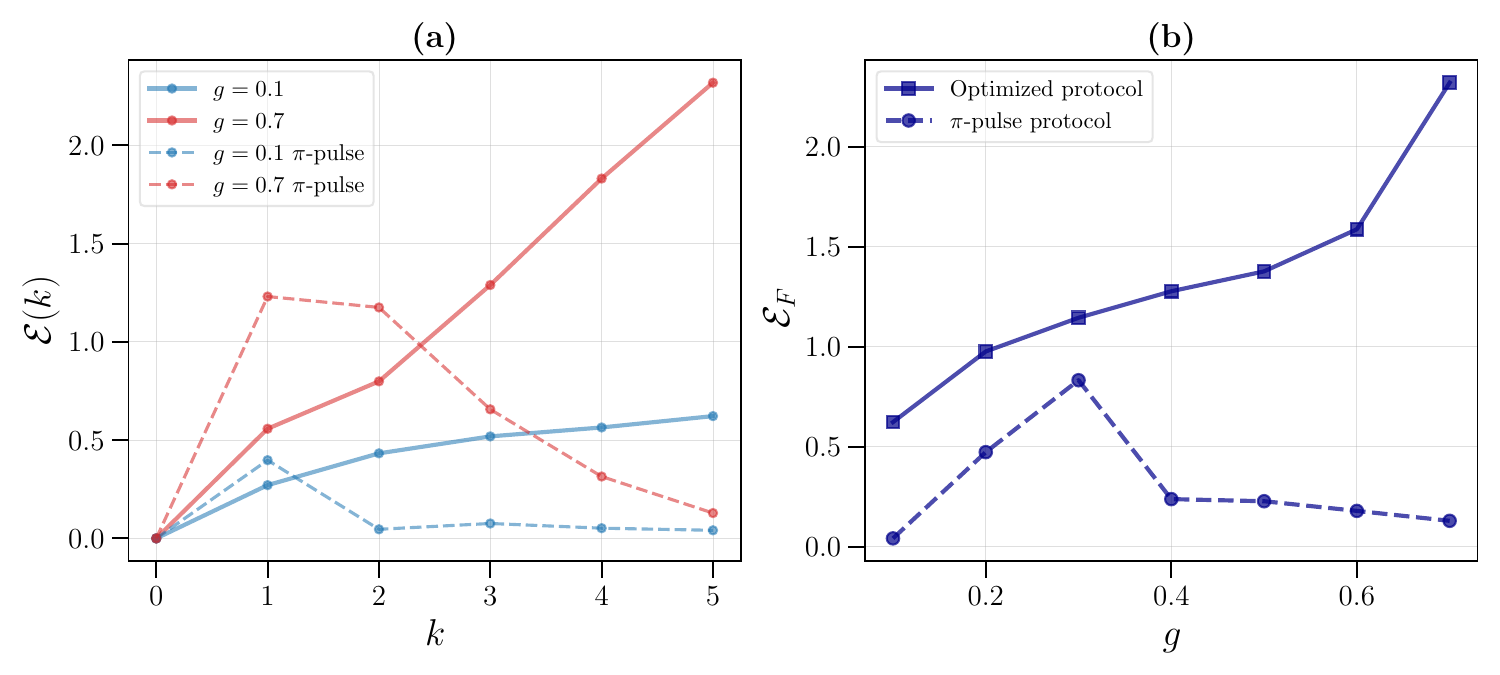} 
    \caption{Comparison of ergotropy of an open quantum battery with optimized and non-optimized protocols.  
    (a) Optimized battery ergotropy as a function of the number of collisions (solid curves), compared with the non-optimized one resulting from the $\pi$-pulse protocol (dashed curves) defined by the Jaynes--Cummings prescription $\tau_k = \pi/(2 g \sqrt{k})$ with $q=0$. Blue curves correspond to $g=0.1$, while red curves correspond to $g=0.7$. The optimized protocol consistently achieves higher final ergotropy than the $\pi$-pulse protocol, demonstrating the effectiveness of the optimization. 
    (b) Battery final ergotropy as a function of the coupling constant $g$, comparing optimized protocols (solid line) with $\pi$-pulse protocols (dashed line). 
    Across all values of $g$, the optimized protocols consistently yield higher final ergotropy.  
    Fixed parameters: $c=1$, $\omega=1$, $\omega_c=3$, $\eta=0.01$, and $\beta=450$.
    }
    \label{fig:combined_ergotropy}
\end{figure*}

Figure \ref{fig:combined_ergotropy}(a) shows the ergotropy of the optimized protocol as a function of the number of collisions $k$, for the cases $g\in\{0.1,0.7\}$. 
The results confirm that the optimization is indeed highly effective: for both values of $g$, the optimized protocol achieves significantly higher ergotropy compared with the $\pi$-pulse protocol ($q=0$ and $\tau_k$ \eqref{eq:tau_pi_pulse}).
At intermediate collisions, the ergotropy in the $\pi$-pulse protocol (dashed lines) can temporarily exceed that in the optimized protocol (solid lines). 
This behavior arises because the optimization is designed to maximize the \emph{final} ergotropy rather than the ergotropy at each intermediate collision.
Overall, jointly optimizing the interaction times and the qubit preparation significantly increases the ergotropy, demonstrating that proper tailoring of the charging protocol can enhance performance.

Figure \ref{fig:combined_ergotropy}(b) shows the final ergotropy of the optimized protocols compared with the $\pi$-pulse protocol, as a function of $g$. This plot summarizes the overall performance of the micromaser battery, demonstrating that the optimized protocols consistently achieve higher final ergotropy than the non-optimized one. In addition, the final ergotropy grows with increasing $g$, reflecting the enhanced energy transfer at stronger coupling strengths.

\begin{figure*}[!t]
    \centering
    \includegraphics[width=1\textwidth]{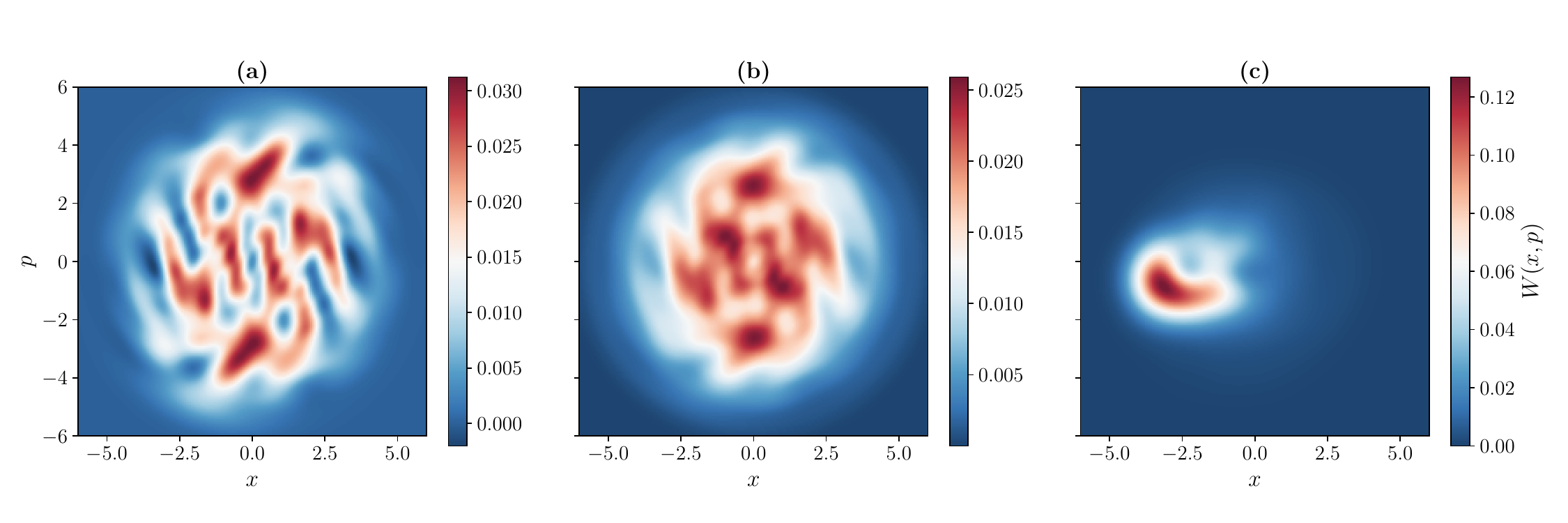}
    \caption{
    Wigner function of the battery state after the charging process for three different protocols. 
    (a) Closed-system $\pi$-pulse protocol with $g=0.7$, $\tau_k = \pi/(2 g \sqrt{k})$, $q=0$, $c=1$, and $\omega=1$. 
    (b) Open-system $\pi$-pulse protocol with the same system parameters and environmental parameters $\omega_c=3$, $\beta=450$, and $\eta=0.01$. 
    (c) Optimized open-system protocol, corresponding to the optimal parameters used for $g=0.7$ in figure~\ref{fig:combined_ergotropy}(a). 
    Panel (a) exhibits small negativities typical of quantum coherence, while panels (b) and (c) show strictly non-negative distributions due to dissipative effects. 
    Panels (a) and (b) display irregular, low-amplitude distributions near the origin, indicative of moderately excited battery states, whereas panel (c) shows a distribution with larger amplitudes displaced away from the origin, signaling a higher energy content and enhanced charging performance.
    }
    \label{fig:wigners}
\end{figure*}

To further illustrate the effectiveness of the optimized protocol, we analyze the Wigner quasi-probability distribution of the battery state after the charging process.  
Unlike classical probability distributions, it can take negative values, reflecting the intrinsically quantum nature of the state. In addition to capturing the state’s phase-space structure, the Wigner function also encodes information about its energy, as expectation values of observables, including the Hamiltonian, can be computed as integrals over phase space weighted by the Wigner function \cite{LopezCarreno2025, Case2008}. 
Figure \ref{fig:wigners} shows the Wigner functions for three cases: the closed-system $\pi$-pulse protocol (figure \ref{fig:wigners}(a)), the open-system $\pi$-pulse protocol (figure \ref{fig:wigners}(b)), and the optimized open-system protocol (figure \ref{fig:wigners}(c)), all corresponding to the parameters used for $g=0.7$ in figure \ref{fig:combined_ergotropy}(a). In the first case, the Wigner function exhibits some negativities, whereas in the other two cases the distributions remain strictly non-negative, likely due to the dissipative nature of the open-system evolution, which tends to suppress quantum features that would otherwise produce negative regions.
Qualitative differences are clearly visible: the non-optimized protocols, i.e., both the closed-system and open-system $\pi$-pulse protocols shown in panels (a) and (b), display an irregular, low-amplitude distribution near the origin, indicative of a moderately excited state. In contrast, the optimized protocol in panel (c) produces a distribution with larger amplitudes displaced away from the origin, signaling a higher energy content. This directly demonstrates that the optimization substantially enhances the charging performance.

\subsubsection{Stabilization}
Charging a battery is only the first step in its operation; the second is stabilizing the injected energy against losses.
After optimizing the charging process, we now turn to the problem of stabilizing the ergotropy stored at the end of the charging stage. 
Our goal is to identify a strategy that maintains the ergotropy of the battery approximately constant over a prescribed time interval.
Indeed, if the stream of qubits is simply switched off and the cavity is allowed to evolve freely, dissipation inevitably causes a progressive loss of energy and, consequently, of ergotropy.
A dedicated stabilization protocol is therefore required to preserve the stored useful work against environmental degradation.
 
\begin{figure*}[!t]
    \centering
    \includegraphics[width=0.8\textwidth]{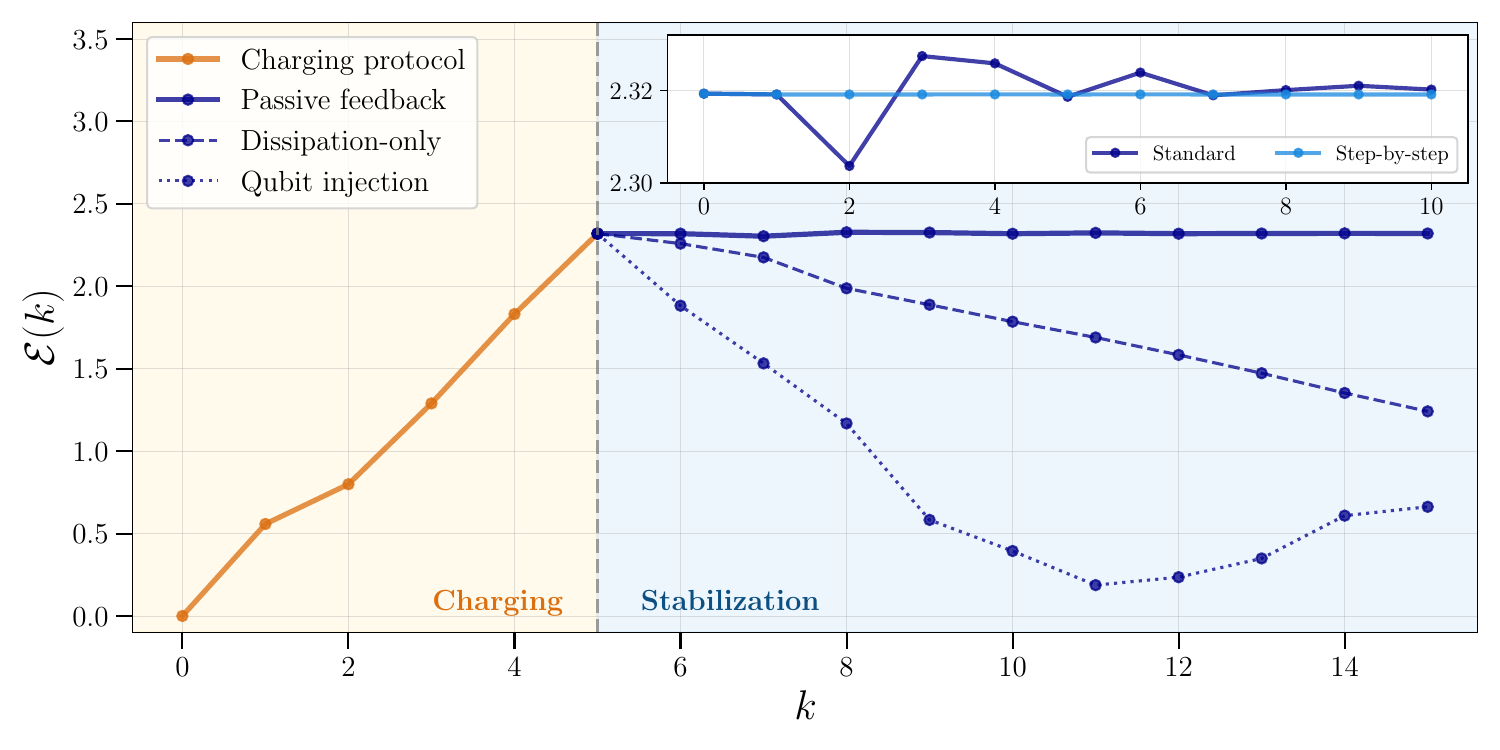}
    \caption{Charging and stabilization of the open quantum battery. 
    The charging protocol corresponds to the optimized process with $g=0.7$ shown in figure~\ref{fig:combined_ergotropy}(a).
    The stabilization stage is compared across three scenarios: the passive-feedback protocol with optimized parameters, the continued injection of qubits without measurement using the same \(q\) and \(\{\tau_k\}\) as in the optimized stabilization protocol, and the evolution under dissipation alone. 
    Parameters are fixed at \(g=0.7\), \(c=1\), \(\omega=1\), \(\eta=0.01\), \(\omega_c=3\), and \(\beta=450\). 
    The passive-feedback protocol successfully maintains the stored ergotropy, while dissipation alone causes gradual decay and qubit injection without measurement leads to a rapid drop in ergotropy after only 10 collisions. 
    The inset shows the optimized passive-feedback stabilization protocol obtained from two different procedures: the standard optimization reported in the main figure and a step-by-step approach in which the interaction times \(\{\tau_k\}\) are sequentially adjusted to stabilize the ergotropy after each collision, while keeping \(q\) fixed at the value obtained from the standard optimization. 
    The step-by-step procedure yields a smoother and more stable ergotropy profile with reduced fluctuations compared to the standard protocol.
    }
    \label{fig:ergotropy_charging_stabilization}
\end{figure*}

As a first approach, we simply leveraged the drive represented by the stream of qubits which sequentially interact with the cavity as a mechanism to counteract dissipation. Even upon optimizing over the parameters $(q,\{\tau_k\})$, this approach fails to stabilize the ergotropy, as no optimal parameters can be identified that preserve the stored ergotropy. Instead, the ergotropy decays even faster than if the cavity were left to evolve under dissipation alone, as shown in figure \ref{fig:ergotropy_charging_stabilization}.
This behavior can likely be attributed to the continued interaction with the qubits, which progressively degrades the purity of the battery.

To improve this scheme, we devised a different strategy, in which each qubit is measured after its collision with the cavity.
We refer to this mechanism as \emph{passive feedback} because the measurement outcomes are not used to actively modify subsequent controls; rather, the measurement backaction itself alters the cavity state, counteracting dissipation and keeping the ergotropy accumulated during the charging stage approximately constant throughout the stabilization interval. To identify the most effective stabilization strategy, we adjusted the same control parameters optimized for the charging process, the qubit population \(q\) and the interaction times \(\{\tau_k\}\), but with a different objective: rather than maximizing the final ergotropy, we now aim to minimize its deviation from the post-charging value, thereby keeping the stored ergotropy approximately constant.

For this purpose, we considered a measurement-based collision model: after each battery-qubit collision, a projective measurement of the qubit energy is performed on the joint battery-qubit state $\rho_{B,q}$. The energy measurement is described by the projectors $\{\hat{\Pi}_x^q = \dyad{x}\}$, satisfying $\sum_x \hat{\Pi}_x^q = \mathbb{I}_{q}$ and $\hat{\Pi}_x^q \hat{\Pi}_{x'}^q = \delta_{x x'} \hat{\Pi}_x^q$, where the two possible outcomes $x\in\{g,e\}$
correspond to the eigenvalues $\mp \omega/2$, respectively. Each sequence of outcomes defines a quantum trajectory $\gamma =(x_1,x_2,\ldots)$, occurring with probability $p_\gamma$.
The necessity of tracking the outcomes can be illustrated by examining the single collision. Upon obtaining outcome $x$, the selective post-measurement state of the battery is $\rho_{B \vert x}'=\Trace_q[\hat{\Pi}_x^q \rho_{B,q} \hat{\Pi}_x^q]/p_x$ with probability $p_x = \Trace[\hat{\Pi}_x^q \rho_{B,q}]$. If the outcome is ignored, the battery is described by the non-selective post-measurement state which is equivalent to the unmeasured state, $\rho_B' = \Trace_q[\sum_{x} \hat{\Pi}_x^q \rho_{B,q} \hat{\Pi}_x^q] = \Trace_q[\rho_{B,q}]=\rho_B$. Therefore, a protocol that discards the measurement outcomes reduces to a sequence of interactions without measurements, i.e., our initial approach we found ineffective at stabilizing the ergotropy. 
Instead, a protocol that keeps track of the measurement outcomes can yield larger ergotropy by exploiting the measurement record to extract work from the selective post-measurement states. In fact, the resulting average ergotropy defines the daemonic ergotropy \cite{Francica2017,bernards2019entropy,Morrone2023,Elyasi2025}
\begin{equation}
\mathcal{E}_D(\rho_B,\{\hat{\Pi}_x^q\}) \equiv \sum_{x}p_x \mathcal{E}(\rho_{B \vert x}') \geq \mathcal{E}(\rho_B),
\end{equation}
which quantifies the maximum work extractable by leveraging information gained from projective measurements on a correlated ancillary system.
Extending this reasoning to a sequence of $N$ collisions followed by selective measurements, our goal is to identify the set of parameters \( (q, \{\tau_k\}) \) that best stabilizes the stored ergotropy by leveraging the qubit measurement record. For a given parameter set, 
we compute the weighted average ergotropy after each collision,
\begin{equation}
\bar{\mathcal{E}}(k) = \sum_{\gamma} p_{\gamma}\, \mathcal{E}_{\gamma}(k),
\label{eq:w_avg_ergotropy}
\end{equation}
where $\mathcal{E}_{\gamma}(k)$ is the ergotropy along the trajectory $\gamma$  after the $k$th collision.
To identify the optimal stabilization strategy, we define a cost function that quantifies the deviation of this average ergotropy from its initial value $\mathcal{E}_\mathrm{in}$ at the beginning of the stabilization stage,
\begin{equation}
\mathcal{C}(q, \{\tau_k\}) = \sum_{k=0}^{N} \big[\, \bar{\mathcal{E}}(k) - \mathcal{E}_\mathrm{in} \,\big]^2,
\label{eq:costfunction}
\end{equation}
where we set $N=10$.
Minimizing $\mathcal{C}$ identifies the control parameters that best preserve the stored ergotropy throughout the stabilization protocol.

Figure~\ref{fig:ergotropy_charging_stabilization} illustrates the charging and stabilization processes of the quantum battery. 
For the stabilization stage, we compare three scenarios: the passive-feedback protocol with optimized parameters, the continuous injection of qubits without measurement using the values of \(q\) and \(\{\tau_k\}\) obtained from the optimization of the stabilization protocol, and the evolution under dissipation alone.
The fixed parameters used in the simulations are \(g=0.7\), \(c=1\), \(\omega=1\), \(\eta=0.01\), \(\omega_c=3\), and \(\beta=450\). 
Thus, the charging process corresponds to the optimized protocol with $g=0.7$ shown in figure~\ref{fig:combined_ergotropy}(a).
The results show that the passive-feedback protocol, combined with optimized control parameters, effectively stabilizes the battery. 
In contrast, free evolution under dissipation leads to a gradual decay of ergotropy, while injecting qubits without measurement performs even worse for the parameters considered, causing a significant drop in ergotropy after only 10 collisions.

In the inset of figure \ref{fig:ergotropy_charging_stabilization}, we compare the performance of the passive-feedback stabilization protocol under two different optimizations: the standard approach described above (the cost function~\eqref{eq:costfunction} is minimized over the whole trajectory), and a step-by-step approach, in which stabilization is enforced at each collision (the cost function $\mathcal{C}(\tau_k) = [\bar{\mathcal{E}}(k) - \mathcal{E}_\mathrm{in}\big]^2$ is minimized at each collision).
In the latter approach, the interaction times $\{\tau_k\}$ are optimized sequentially after each collision, 
while keeping $q$ fixed at the value obtained from the standard optimization, so as to maintain the ergotropy approximately constant throughout the evolution.
This procedure results in a smoother ergotropy profile with reduced fluctuations compared to the standard protocol. 

We point out that, while step-by-step optimization is meaningful in the stabilization protocol, it is not suitable in the charging protocol, where the goal is to maximize the \textit{final} ergotropy $\mathcal{E}_F$. Step-by-step optimization is a ``greedy'' strategy that locally optimizes each collision, and a trajectory through locally optimal solutions may fail to reach the global optimum. In our setting, step-by-step maximization of the ergotropy $\mathcal{E}(k)$ at each collision does not reproduce the globally optimized protocol for $\mathcal{E}_F$, typically resulting in lower final ergotropy and worse overall performance.
In contrast, in the stabilization protocol the objective is to maintain a nearly constant ergotropy throughout the evolution, $\bar{\mathcal{E}}(k)\approx \mathcal{E}_{\rm in}$.
In this case, a step-by-step optimization, in which ergotropy fluctuations are minimized at each collision, is more appropriate and numerically accurate than a global optimization, in which ergotropy fluctuations are minimized over the entire trajectory.
Indeed, we have numerically identified a stabilization protocol that maintains constant ergotropy (see inset of figure~\ref{fig:ergotropy_charging_stabilization}). Although this protocol is, in principle, a solution of both the global and the step-by-step optimization, the latter optimization proves more effective in practice at identifying it, thereby achieving a more accurate stabilization of the stored ergotropy.

In conclusion, our measurement-based passive-feedback protocol relies on the selective backaction of projective energy measurement on the ancillary qubit, highlighting that stabilization is not merely a byproduct of the competition between the drive (sequential interactions) and dissipation, but is fundamentally driven by information. Our decision to restrict this strategy to the stabilization stage, despite its potential to enhance the charging as well, is motivated by the inherent energetic and entropic costs of measurements.
While projective energy measurements on a total system have a vanishing energetic cost on average \cite{Gherardini2020}, the average energetic cost of projectively measuring the energy of a correlated, interacting ancillary system is finite
\begin{align}
    \langle \Delta E_{\rm meas}\rangle  
    & = \sum_x p_x \Tr[\hat{H}_{B,q}(\rho_{B,q \vert x}' - \rho_{B,q})]\nonumber\\
    & = \Tr[\hat{V}_{B,q} (\rho_{B,q}' - \rho_{B,q})],
\end{align}
where $\rho_{B,q \vert x}' = \hat{\Pi}_x^q \rho_{B,q} \hat{\Pi}_x^q/p_x$ 
with $p_x = \Trace[\hat{\Pi}_x^q \rho_{B,q}]$ and $\rho_{B,q}' = \sum_x \hat{\Pi}_x^q \rho_{B,q} \hat{\Pi}_x^q$.  
For arbitrary states $\rho_{B,q}$, the cost is in general non-zero, since $[\hat{V}_{B,q},\hat{\Pi}_x^q] \neq 0$. 
Beyond the energetic cost, one must also account for the entropic cost associated with the erasure of the measurement record (memory). Formally, these processes fall under the framework of quantum feedback control and erasure protocols, where the total cost of measurement and memory reset is governed by the second law of information thermodynamics \cite{Minagawa2025npj}.
Although this measurement scheme can, in principle, enhance the charging process as well, the latter remains effective even in the absence of measurements. Consequently, by employing measurements only where they are strictly required for stabilization, we reduce the total cost of the battery's operation.

\section{Conclusion}
\label{sec:conclusion}
We have investigated a dissipative micromaser quantum battery in the ultrastrong coupling (USC) regime, focusing on (i) the optimization of the charging protocol and (ii) the stabilization of the stored ergotropy against losses. A single-mode electromagnetic cavity, acting as battery, repeatedly interacts with a stream of qubits, acting as chargers.
Because counter-rotating terms cannot be neglected in the USC regime, each qubit-cavity interaction is governed by the full Rabi Hamiltonian.
Assuming weak coupling to a Markovian environment, we have numerically simulated the dissipative dynamics of the micromaser battery using a GKLS master equation.
Crucially, dissipation does more than just provide a realistic physical picture: it actively stabilizes the charging process by preventing the unbounded energy growth characteristic of unitary evolution, and drives the battery toward a steady state with finite ergotropy and higher purity than the corresponding non-dissipative dynamics.
Ultimately, this reveals the micromaser battery's capability to store useful energy despite environmental losses. 

(i) We numerically optimized the battery's charging protocol over the interaction times $\{\tau_k\}$ and qubit population parameter $q$ to maximize the final ergotropy using a limited number of chargers.
The optimization significantly enhances the battery's performance: after only a few collisions, the stored ergotropy is consistently higher than that of benchmark strategies, such as the $\pi$--pulse protocol representing the optimal charging protocol in the weak-coupling limit (Jaynes--Cummings model). Furthermore, our results show that stronger light-matter couplings within the USC regime lead to faster battery charging while yielding greater stored ergotropy. 

(ii) To stabilize the battery's ergotropy after the charging stage, we introduced a measurement-based passive-feedback protocol where each charger qubit is measured after interacting with the cavity. By numerically optimizing $q$ and $\{\tau_k\}$ to minimize deviations from the value of the post-charging ergotropy, this mechanism effectively
counteracts dissipative losses and maintains the stored ergotropy approximately constant over a desired time interval. This passive-feedback protocol remarkably outperforms other stabilization strategies we considered; notably, driving the battery with unmeasured qubits can degrade the stored ergotropy faster than bare dissipation.

We emphasize that our open-system simulations rely on conservative assumptions for dissipation, with a dimensionless single-photon loss rate $\gamma/\omega \simeq 0.045$ (see Sec.~\ref{sec:open_dynamics}). This value exceeds by approximately two orders of magnitude the typical cavity decay rates $\gamma/\omega \sim 10^{-4}$ reported in circuit-QED experiments in the USC regime \cite{Niemczyk2010,Beaudoin2011,Fink2008,Grimsmo2013}. As a consequence, our results are likely to underestimate the achievable performance in realistic experimental platforms.

In conclusion, our work reveals the rich interplay between strong light-matter interaction, environmental dissipation, and optimal-control strategies in quantum batteries. We have shown that a dissipative micromaser quantum battery operating in the USC regime can be effectively charged and stabilized, and its performance significantly enhanced through optimal control techniques. This paves the way for designing  robust quantum energy-storage devices that leverage dissipation and control as complementary resources, and 
several promising extensions of this work can be envisaged.
First, a more realistic characterization of the protocol should account for time constraints, as they are not necessarily captured by fixing the number of collisions. In this context, multiobjective optimizations aimed at balancing charging power (stored ergotropy per unit time) and the final stored ergotropy provide a more appropriate framework.
Furthermore, a comprehensive thermodynamic assessment of the protocol's operational costs remains necessary and subject for future study; this includes evaluating the minimum control power, the relative
(excess) stabilization cost \cite{Gherardini2020}, the costs associated with nonideal projective measurements \cite{Guryanova2020quantum}, measurement and information erasure \cite{Minagawa2025npj}, and cooling the apparatus \cite{Taranto2023prxq}.
Finally, exploring non-Markovian dynamics through initially correlated or pairwise interacting charger qubits, the simultaneous injection of multiple qubits into the cavity, continuous monitoring of the environment \cite{WisemanMilburn2009,AlbarelliGenoni2024,Mitchison2021, YaoShao2022, deOliveira2025, Francica2017, Manzano2018, Morrone2023, cenedese2025arxiv_qb} or bath engineering could open further possibilities for control and work extraction.

\begin{acknowledgments}
G.B. acknowledges useful discussions with Dario Ferraro.
M.C. and G.B. acknowledge support from INFN through the project ``QUANTUM'' and from the European Union-NextGenerationEU through the ``Solid State Quantum Batteries: Characterization and Optimization'' (SoS-QuBa) project (Prot. 2022XK5CPX), in the framework of the PRIN 2022 initiative of the Italian Ministry of University (MUR) for the National Research Program (PNR). This project has been funded within the programme ``PNRR Missione 4- Componente 2 Investimento 1.1 Fondo per il Programma Nazionale di Ricerca e Progetti di Rilevante Interesse Nazionale (PRIN)'', funded by the European Union-Next Generation EU.
L.R. acknowledges support from University of Pavia through the project 
``Termodinamica di precisione per sistemi aperti quantistici'',
funded within the ``Fondo Ricerca e Giovani 2024'' programme, and from INFN through the project ``BELL''.
L.G. acknowledges support from the PNRR MUR project PE0000023-NQSTI, ``National Quantum Science and Technology Institute''.
G.A.F. acknowledges support from the ICSC - Centro Nazionale di Ricerca in High-Performance Computing, Big Data and Quantum Computing. L.G. and G.A.F. acknowledge support from PRIN 2022 ``SuperNISQ'' and from 
the University of Catania, Piano Incentivi Ricerca di Ateneo 2024-26, project TCMQI.
\end{acknowledgments}

\bibliography{references}

\end{document}